\documentclass[11pt]{article}
\usepackage[letterpaper, left=1in, top=1in, right=1in, bottom=1in,nohead,includefoot]{geometry}
\usepackage{color,charter,graphicx,natbib,here,setspace,multirow,amsmath,amssymb,amsfonts,enumitem,placeins,amsthm,cancel,mathrsfs}

\usepackage[dvipsnames]{xcolor}
	\usepackage[pdftex]{hyperref}
	\hypersetup{colorlinks,
	citecolor=NavyBlue  
	}

\bibpunct{(}{)}{;}{a}{}{,} 

\def\figdir{figurescolorjpg} 

\def\x{\mbox{\boldmath$x$}}

\def\btheta{\mbox{\boldmath$\theta$}}

\def\mM{\mathcal{M}}

\theoremstyle{definition}

\newtheorem{theorem}{Theorem}[section]

\newtheorem{proposition}{Proposition}[section]

\begin{document}
\title{Mean-shift least squares model averaging}  
\author{Kenichiro McAlinn\thanks{Department of Statistical Science, Fox School of Business, Temple University,  Philadelphia, PA 19122. {\scriptsize  Email: kenichiro.mcalinn@temple.edu} }  
\,  \& K\={o}saku Takanashi\thanks{RIKEN Center for Advanced Intelligence Project, Tokyo, Japan. {\scriptsize  Email: kosaku.takanashi@riken.jp } }
}
 
\maketitle\thispagestyle{empty}\setcounter{page}0

\begin{abstract}
 
 This paper proposes a new estimator for selecting weights to average over least squares estimates obtained from a set of models.
 Our proposed estimator builds on the Mallows model average (MMA) estimator of \cite{hansen2007least}, but, unlike MMA, simultaneously controls for location bias and regression error through a common constant.
 We show that our proposed estimator-- the mean-shift Mallows model average (MSA) estimator-- is asymptotically optimal to the original MMA estimator in terms of mean squared error.
A simulation study is presented, where we show that our proposed estimator uniformly outperforms the MMA estimator.

\bigskip
\noindent
{\em Keywords}: Model averaging, Mallows criterion, series estimators, optimality.
\end{abstract}

\newpage

\section{Introduction \label{sec:intro}}
The goal of model averaging, where several models are averaged over, is to reduce estimation variance while controlling for bias arising from omitted variables (regression error).
 \cite{hansen2007least} proposed the Mallows model averaging (MMA) estimator, which is shown to be asymptotically optimal with regard to the fitted estimates  achieving the minimum squared error in a class of discrete model averaging estimators.
 However, while linear averaging (including MMA) can minimize regression error, it cannot simultaneously reduce regression error and location bias; the bias arising from non-zero expectation in the misspecification bias.
 In this paper, we propose a novel model averaging estimator that controls for location bias while reducing regression error.
  We call this new estimator the mean-shift Mallows model averaging (MSA) estimator.
 We relax the condition of discrete estimators  to continuous estimators, similar to that of \cite{wan2010least}, though under weaker conditions.
 We show that the proposed MSA estimator is asymptotically optimal to the MMA estimator.
 
 Model averaging (or forecast combination/ensemble learning) has been a staple and necessary tool in the field of econometrics, statistics, and machine learning.
Since the seminar paper by \cite{BatesGranger1969}, the potency of model averaging has been proven over multiple applications and contexts.
Recently, the field has received a surge of interest across many disciplines due to an increase in usage and availability of full forecast densities arising from more complex models.
In the statistical literature, there is a large literature on model averaging, particularly for Bayesian methods.
 For example, Bayesian model averaging \citep[BMA:][]{raftery1997bayesian,hoeting1999bayesian} has become standard for Bayesians dealing with model uncertainty.
 In machine learning, ensemble methods, including bagging \citep{breiman1996bagging}, boosting \citep{schapire2003boosting}, and stacking \citep{dvzeroski2004combining}, which are all linear model averaging (often equal weight), have been used extensively to alleviate overfitting.
 In econometrics, increased usage in density forecasts has stimulated the field for more advanced strategies for model combination \citep[e.g.][]{Terui2002,HallMitchell2007,Amisano2007,Hooger2010,Kascha2010,Geweke2011,Geweke2012,Billio2012,Billio2013,Aastveit2014,Fawcett2014,Pettenuzzo2015,Negro2016,yao2018using,aastveit2018evolution,Aastveit2015}, with success in applications across many subdisciplines of economics.

While practical success of model averaging is abundant, theoretical considerations, in particular with regard to estimation of averaging weights, have been somewhat limiting.
In terms of theoretical results for averaging models over model selection, \cite{BatesGranger1969} show that a combination of two forecasts can yield improved forecasts over the two individual forecasts, in terms of mean squared forecast error.
In the machine learning literature, \cite{louppe2014understanding} show that, for averaging regression trees (i.e. random forests), ensemble learning lowers variance while maintaining bias.
In the Bayesian literature, when the true model is nested in the candidate models ($\mM$-closed, as it is referred, following \citealt{bernardo2009bayesian}), BMA is known to asymptotically converge to the true model.
However, under an $\mM$-open setting, where the true model is not nested, BMA tends to converge to the ``wrong" model, which is often not even the ``best" model.

In terms of estimating the weights for model averaging, \cite{hansen2007least} showed that minimizing the Mallows criterion is asymptotically optimal in a class of discrete model average estimators in terms of mean squared error.
\cite{wan2010least} relaxes this assumption of discrete estimators to continuous estimators, and show that the optimality holds.
Further theoretical results are developed for more complex settings \citep[see, e.g.,][]{hansen2012jackknife,zhang2013model}.
While the MMA estimator is optimal for model averaging, it cannot simultaneously minimize the bias arising from omitted variables and location bias, due to its limited parameter dimension.

 Building on the MMA estimator in \cite{hansen2007least}, we propose a new model averaging estimator that controls for both biases: the mean-shift Mallows model averaging estimator (MSA).
Our main contribution is to show that our proposed MSA estimator is asymptotically optimal to the MMA estimator, achieving lower risk.
While our MSA estimator is frequentist by nature, our motivation is, nonetheless, Bayesian.
In particular, our development is inspired by the recent practical and theoretical success of Bayesian predictive synthesis (BPS); a coherent Bayesian framework to combine information-- in particular, density forecasts-- from multiple sources \citep[see, e.g.,][]{mcalinn2019dynamic,mcalinn2017multivariate,takanashi2019predictive}.
Part of the success of BPS is in considering forecasts as ``data" to be used in prior-posterior updating, treating them as latent factors.
The flexibility of this framework, of which MMA is a special case, has lead to development of more elaborate model averaging approaches.
One of which is to control for location bias arising from misspecification by incorporating a common constant, which shifts the mean to counteract the bias.
This idea is central to our proposed MSA estimator.

Section.~\ref{sec:setup} discusses the setup as well as review the results from \cite{hansen2007least}.
Section.~\ref{sec:MSA} introduces the mean-shift Mallows model averaging estimator and its sampling properties.
Section.~\ref{sec:sim} provides finite sample evidence in favor of the proposed MSA estimator.

\section{Model averaging under misspecification \label{sec:setup}}
Let $\left\{y_{i}\right\} _{i\in n}\in\mathbb{R}$ and $\x$ be a countably infinite
vector.
The data generating process \citep[DGP: following][]{hansen2007least} is
\begin{align}\label{dgp}
y_{i}=\sum_{j=1}^{k_{m}}\theta_{j}x_{ji}+\sum_{j=k_{m}+1}^{\infty}\theta_{j}x_{ji}+\varepsilon_{i} \quad \varepsilon_{i}\sim N(0,\sigma^2).
\end{align}
Note that the linearity of the DGP is not restrictive, as it includes series expansions.
Consider each model, $m=1,2,...,$ uses the first $k_{m}$ elements of  $x_{ji}$, where $0<k_1<k_2<...$, to construct an approximating model.
This implies that the models are ordered, though this is not problematic when there are series expansions.

The misspecification bias (regression error) is
\begin{align}
b_{mi}=\sum_{j=k_{m}+1}^{\infty}\theta_{j}x_{ji},
\end{align}
which is the component in the DGP that is never captured by any of the models, thus, under this setup, all models are misspecified.
The predictive mean is,
\begin{align}
\hat{\mu}_{m}=X_{m}\hat{\btheta}_{m}=P_{m}Y=X_{m}\left(X_{m}^{\top}X_{m}\right)^{-1}X_{m}^{\top}Y,	
\end{align}
where $P_{m}=X_{m}\left(X_{m}^{\top}X_{m}\right)^{-1}X_{m}^{\top}$ is the projection matrix and $\hat{\btheta}_{m}$ is the vector of least squares estimates.
The full model, $M=M_{n}\leqq n$, is an integer for which $X_{k_{M}}^{\top}X_{k_{M}}$ is invertible.

The weights for model averaging is specified as
\begin{align}
H_{n}=\left\{ W\in\left[0,1\right]^{M}:\sum_{m=1}^{M}w_{m}=1\right\},	
\end{align}
with the resulting linear averaged predictive mean being
\begin{align}
\hat{\mu}\left(W\right)=P\left(W\right)Y=\sum_{m=1}^{m}w_{m}P_{m}Y.	
\end{align}
Then, we have the following properties:
\begin{subequations}
\begin{align}
	P\left(W\right)&=\sum_{m=1}^{m}w_{m}P_{m},\\
	P\left(W\right) &\textrm{ is symmetric but not idempotent: } P^{\top}\left(W\right)=P\left(W\right),\quad P^{2}\left(W\right)\neq P\left(W\right)\\
	\textrm{tr}\left\{ P\left(W\right)\right\} &=\sum_{m=1}^{m}w_{m}k_{m}\\
	\textrm{tr}\left\{ P\left(W\right)P\left(W\right)\right\} &=\sum_{m=1}^{m}\sum_{l=1}^{m}w_{m}w_{l}\min\left(k_{l},k_{m}\right)=W^{\top}\varGamma_{m}W\\
	\textrm{tr}\left\{ P_{m}\right\}  & =k_{m}\\
\textrm{tr}\left\{ P_{m}P_{\ell}\right\}  & =\textrm{tr}\left\{ P_{\min\left(k_{\ell},k_{m}\right)}\right\} =\min\left(k_{\ell},k_{m}\right)\\
\textrm{tr}\left\{ P\left(W\right)\right\}  & =\textrm{tr}\left\{ \sum_{m=1}^{m}w_{m}P_{m}\right\} =\sum_{m=1}^{m}w_{m}k_{m}=k\left(W\right)\\
\textrm{tr}\left\{ P\left(W\right)P\left(W\right)\right\}  & =\sum_{m=1}^{m}\sum_{\ell=1}^{m}w_{m}w_{\ell}\min\left(k_{\ell},k_{m}\right)=W^{\top}\varGamma_{m}W.
\end{align}
\end{subequations} 

Define the risk (i.e. the conditional squared error), $R\left(W\right)=\mathbb{E}\left[L_{n}\left(W\right)\left|X\right.\right]$, as
\begin{alignat*}{1}
 & R\left(W\right)\\
= & \left[\begin{array}{c}
w_{1}\\
w_{2}\\
w_{3}\\
\vdots\\
w_{m}
\end{array}\right]^{\top}\left(\underbrace{\left[\begin{array}{ccccc}
a_{1} & a_{2} & a_{3} & \cdots & a_{m}\\
a_{2} & a_{2} & a_{3} & \cdots & a_{m}\\
a_{3} & a_{3} & a_{3} & \cdots & a_{m}\\
\vdots & \vdots & \vdots & \ddots & a_{m}\\
a_{m} & a_{m} & a_{m} & \cdots & a_{m}
\end{array}\right]}_{A_{m}}+\sigma^{2}\underbrace{\left[\begin{array}{ccccc}
k_{1} & k_{1} & k_{1} & \cdots & k_{1}\\
k_{1} & k_{2} & k_{2} & \cdots & k_{2}\\
k_{1} & k_{2} & k_{3} & \cdots & k_{3}\\
\vdots & \vdots & \vdots & \ddots & \vdots\\
k_{1} & k_{2} & k_{3} & \cdots & k_{m}
\end{array}\right]}_{\varGamma_{m}}\right)\left[\begin{array}{c}
w_{1}\\
w_{2}\\
w_{3}\\
\vdots\\
w_{m}
\end{array}\right]\\
 & =W^{\top}\left(A_{m}+\sigma^{2}\varGamma_{m}\right)W,
 \end{alignat*}
 where
\begin{align*}
A_{m}&+\sigma^{2}\varGamma_{m}>0,\textrm{ if }a_{1}>0\\
a_{m}&=b_{m}^{\top}\left(I-P_{m}\right)b_{m},
\end{align*}
following \cite{hansen2007least} (Lemma 2).
This shows that the risk is a quadratic function with regard to the weight vector, $W$.
From this, we have
\begin{align*}
\mathbb{E}\left[L_{n}\left(W\right)\left|X\right.\right] & =\mu^{\top}\left(I-P\left(W\right)\right)\left(I-P\left(W\right)\right)\mu+W^{\top}\sigma\varGamma_{M}W.
\end{align*}

The residual from the linear projection can be written as
\begin{align*}
\left(I-P_{m}\right)\mu & =\left(I-P_{m}\right)\left(\sum_{j=1}^{m}\theta_{j}x_{j}+\sum_{j=m+1}^{\infty}\theta_{j}x_{j}\right)=\left(I-P_{m}\right)b_{m}\\
\left(I-P\left(W\right)\right)\mu & =\sum_{m=1}^{M}w_{m}\left(I-P_{m}\right)b_{m}=B_{n}W\\
\mu\left(I-P\left(W\right)\right)\left(I-P\left(W\right)\right)\mu & =W^{\top}B_{n}^{\top}B_{n}W=W^{\top}A_{n}W.
\end{align*}
When $\ell\leqq m$, the residual is
\begin{alignat*}{1}
P_{\ell}P_{m} & =P_{\ell}\\
\left(I-P_{m}\right)b_{\ell} & =\left(I-P_{m}\right)b_{m}.
\end{alignat*}


The Mallows criterion for the model average estimator is
\begin{alignat*}{1}
C_{n}\left(W\right) & =W^{\top}\left(y-x\hat{\beta}\right)^{\top}\left(y-x\hat{\beta}\right)W+2\sigma^{2}K^{\top}W,
\end{alignat*}
where 
\begin{align*}
K=\left(k_{1},k_{2},\cdots,k_{m}\right)^{\top},
\end{align*}
 is the vector of model dimensions and the collection of residuals is
\begin{align*}
\left(y-x\hat{\beta}\right) & =\left(\hat{\varepsilon}_{1},\cdots,\hat{\varepsilon}_{M}\right)_{n\times M}.
\end{align*}
For the MMA estimator, the weight vector is selected by minimizing the Mallows criterion.
This has the expectation \citep[][Lemma 3]{hansen2007least},
\begin{align*}
\mathbb{E}\left[C_{n}\left(W\right)\right] & =\mathbb{E}\left[L_{n}\left(W\right)\right]+n\sigma^{2}.
\end{align*}

The empirical weight vector is
\[
\hat{W}_{n}=\arg\min_{W}C_{n}\left(W\right).
\]
for which \cite{hansen2007least} showed that this criterion is optimal, as we summarize below.

Limiting the weight vector to a finite $N$ number of discrete values, where the possible values, $W\in\left[0,1\right]$, are the $N+1$, 
\[
w_{m}\in\left\{ 0,\frac{1}{N},\frac{2}{N},\cdots,1\right\}. 
\]
The set of possible discrete weight vectors is denoted as $H_{n}\left(N\right)$.
When $n\rightarrow\infty$, assume
\[
\xi_{n}=\inf_{W\in H_{n}}R_{n}\left(W\right)\rightarrow\infty
\]
almost surely.
Additionally, assume
\[
\mathbb{E}\left[\left|e_{i}\right|^{4\left(N+1\right)}\left|x_{i}\right.\right]\leqq\kappa<\infty.
\]
Then, the following holds \citep[][Theorem 1]{hansen2007least}:
\[
\frac{L_{n}\left(\hat{W}_{N}\right)}{{\displaystyle \inf_{W\in H_{n}\left(W\right)}}L_{n}\left(W\right)}\rightarrow1.
\]

The theoretical results from \cite{hansen2007least} shows that choosing $\hat{W}$ by minimizing $C_{n}\left(W\right)$ minimizes the loss (but not the risk), as well as achieve the optimal rate.
However, limiting the weight vector to a finite $N$ number of discrete values is somewhat restrictive, which we will later relax.

While the MMA estimator is optimal for a class of discrete model averaging estimators, it does not mean that there is no room for improvement.
This can be easily seen when we decompose the loss as
\begin{alignat*}{1}
\left(\mu\left(W\right)-\mu\right)^{\top}\left(\mu\left(W\right)-\mu\right) & =\left(\mu\left(W\right)-\mathbb{E}\left[y\left|\mathcal{F}_{t}\right.\right]+\mathbb{E}\left[y\left|\mathcal{F}_{t}\right.\right]-\mu\right)^{\top}\left(\mu\left(W\right)-\mathbb{E}\left[y\left|\mathcal{F}_{t}\right.\right]+\mathbb{E}\left[y\left|\mathcal{F}_{t}\right.\right]-\mu\right)\\
 & =\left(\mu\left(W\right)-\mathbb{E}\left[y\left|\mathcal{F}_{t}\right.\right]\right)^{2}+\left(\mathbb{E}\left[y\left|\mathcal{F}_{t}\right.\right]-\mu\right)^{2}+2\left(\mu\left(W\right)-\mathbb{E}\left[y\left|\mathcal{F}_{t}\right.\right]\right)\left(\mathbb{E}\left[y\left|\mathcal{F}_{t}\right.\right]-\mu\right)\\
 & =\left(\mu\left(W\right)-\mathbb{E}\left[y\left|\mathcal{F}_{t}\right.\right]\right)^{2}+\left(\mathbb{E}\left[y\left|\mathcal{F}_{t}\right.\right]-\mu\right)^{2}.
\end{alignat*}
Since the linear projection, $P_{m}\mu$, will ultimately not be the conditional expectation of $y$ given $X_{m}$ and the averaged $P\left(W\right)\mu$ will not be the conditional expectation of $y$ given all the available information, this will leave $\left(\mu\left(W\right)-\mathbb{E}\left[y\left|\mathcal{F}_{t}\right.\right]\right)^{2}$ to be improved.

This improvement can be expound further by decomposing the DGP and identifying the source of possible improvement.
Consider decomposing the DGP into the full model, $M$, and its misspecification bias;
\[
y_{i}=\sum_{j=1}^{k_{M}}\theta_{j}x_{ji}+\sum_{j=k_{M}+1}^{\infty}\theta_{j}x_{ji}+\varepsilon_{i}.
\]
Here, the covariates for $M$, $\left\{ x_{j}\right\} _{j=1,\cdots,K_{M}}$, and the covariates for misspecification bias, $\left\{ x_{j}\right\} _{j=K_{M},\cdots,\infty}$, are considered independent.
The conditional expectation of the full model for $y_{i}$, given all observable covariates, $\left\{ x_{j}\right\} _{j=1,\cdots,k_{M}}$, is
\[
\mathbb{E}\left[y\left|\left\{ x_{j}\right\} _{j=1,\cdots,k_{M}}\right.\right]=\sum_{j=1}^{k_{M}}\theta_{j}x_{ji}.
\]
The conditional expectation  of the misspecification bias of $M$, due to independence, is 
\[
\mathbb{E}\left[b_{Mi}\right]=\mathbb{E}\left[\sum_{j=k_{M}+1}^{\infty}\theta_{j}x_{ji}\right].
\]

If each $\left\{ x_{j}\right\} _{j=1,\cdots,\infty}$ are i.i.d., then $\mathbb{E}\left[b_{Mi}\right]$ does not depend on the sampling order $i$, thus we denote it as $\alpha_{M}$.
The best predictive value, in terms of MSE, can be written as,
\[
\alpha_{M}+\sum_{j=1}^{k_{M}}\theta_{j}x_{ji},
\]
which is to say that the DGP can be decomposed into its location bias, $\alpha_{M}$, and the optimal regression, $\sum_{j=1}^{k_{M}}\theta_{j}x_{ji}$.
This can be written in matrix form (with elements corresponding to $i$) as
\[
\alpha_{M}\mathbf{1}+X_{M}\theta_{M}.
\]
Note that $\mathbf{1}$ is a vector of ones.

The expectation of the misspecification bias for each model, $h$, using $\alpha_{h}$, is
\[
\alpha_{h}=\mathbb{E}\left[b_{i,h}\right]=\mathbb{E}\left[\sum_{j=k_{h}+1}^{\infty}\theta_{j}x_{ji}\right].
\]
The intercept, $\hat{\alpha}_{h}$, is the estimate for $\alpha_{h}$ for each model, $h$.
The predictive value, $\hat{\mu}_{h}$, for each model is,
\[
\hat{\mu}_{hi}=\hat{\alpha}_{h}+\sum_{j=1}^{k_{h}}\hat{\beta}_{j}x_{ji},
\]
where we separate the intercept.
The matrix form (with elements corresponding to $i$) is
\[
\hat{\mu}_{h}=\hat{\alpha}_{h}\mathbf{1}+X_{h}^{-\alpha_{h}}\hat{\boldsymbol{\beta}}_{h}^{-\alpha_{h}}.
\]
Note that $\mathbf{1}$ is a vector of ones, and $X_{h}^{-\alpha_{h}},\hat{\boldsymbol{\beta}}_{h}^{-\alpha_{h}}$ is $X_{h}\hat{\boldsymbol{\beta}}_{h}$ minus the intercept.

Given the above, linear model averaging (including MMA) can be written as
\[
\sum_{h=1}^{k_{M}}w_{h}\left(\hat{\alpha}_{h}\mathbf{1}+X_{h}^{-\alpha_{h}}\hat{\theta}_{h}^{-\alpha_{h}}\right).
\]
The weights, $w_{h}$ are chosen to be as close to the best predictive value of $y$; $\alpha_{M}+\sum_{j=1}^{k_{M}}\theta_{j}x_{ji}$.
The optimal weight vector is $w_{h}^{*}$.
Then, except in rare situations, we have,
\begin{align*}
\alpha_{M} & \neq\sum_{h=1}^{k_{M}}w_{h}^{*}\hat{\alpha}_{h}\\
\sum_{j=1}^{k_{M}}\theta_{j}x_{ji} & \neq\sum_{h=1}^{k_{M}}w_{h}^{*}\left(\sum_{j=1}^{k_{h}}\hat{\theta}_{j}x_{ji}\right).
\end{align*}
Since the weights, $w_{h}$, must be selected by simultaneously minimizing the location bias and regression error, the weights on the location bias need not be zero.
There might exist a vector of weights, $w$, that satisfies $\alpha_{M}=\sum_{h=1}^{k_{M}}w_{h}\hat{\alpha}_{h}$, but that will likely worsen the regression error.
This is due to the dimension for linear averaging missing one dimension with regard to optimizing the averaging weights.
Thus, under linear averaging, there is always a tradeoff between the location bias and regression error, where minimizing one leads to an increase in the other, making all linear averaging estimators (MMA included) suboptimal.

\section{Mean-shift least squares averaging \label{sec:MSA}}
To simultaneously minimize the location bias and regression error, we propose a new estimator that includes a common constant (i.e. an intercept) to the set of models, which we call the mean-shift Mallows model averaging (MSA) estimator.
Our MSA estimator has the following predictive mean:
\begin{align}
\hat{\mu}\left(W,\alpha\right)=\alpha+\sum_{m=1}^{m}w_{m}P_{m}Y,	
\end{align}
where the empirical weights, $\hat{W}$, and intercept, $\hat{\alpha}$, are estimated by minimizing the criterion we develop in this section. 

When we include a common constant, the residual of the linear projection is
\begin{align*}
\left(I-P\left(W\right)\right)\mu-\alpha1 & =\sum_{m=1}^{M}w_{m}\left\{ \left(I-P_{m}\right)b_{m}-\alpha1\right\} =\left(B_{n}-\alpha_{n\times M}\right)W\\
\alpha_{n\times M} & =\left[\begin{array}{cccc}
\alpha & \alpha & \cdots & \alpha\\
\alpha & \ddots &  & \vdots\\
\vdots\\
\alpha & \cdots &  & \alpha
\end{array}\right]
\end{align*}
where
\begin{align*}
W\left(B_{n}-\alpha_{n\times M}\right)\left(B_{n}-\alpha_{n\times M}\right)W &= W^{\top}B_{n}^{\top}B_{n}W\\
 & +\underline{W^{\top}\alpha_{n\times M}^{\top}\alpha_{n\times M}W-W^{\top}B_{n}^{\top}\alpha_{n\times M}W-W^{\top}\alpha_{n\times M}^{\top}B_{n}W}_{\left(1\right)}.
\end{align*}

For the extra term $\left(1\right)$, arbitrarily fixing the weight vector, $W$, we have a quadratic function with regard to $\alpha$:
\begin{alignat*}{1}
 & \alpha^{2}nW^{\top}\left[\begin{array}{cccc}
1 & 1 & \cdots & 1\\
1 & \ddots &  & \vdots\\
\vdots\\
1 & \cdots &  & 1
\end{array}\right]_{M\times M}W-\alpha W^{\top}\left[\begin{array}{cccc}
\sum b_{1} & \sum b_{1} & \cdots & \sum b_{1}\\
\sum b_{2} & \sum b_{2} & \cdots & \sum b_{2}\\
\vdots & \vdots & \vdots & \vdots\\
\sum b_{M} & \sum b_{M} & \cdots & \sum b_{M}
\end{array}\right]_{M\times M}W\\
 & -\alpha W^{\top}\left[\begin{array}{cccc}
\sum b_{1} & \sum b_{2} & \cdots & \sum b_{M}\\
\sum b_{1} & \sum b_{2} & \cdots & \sum b_{M}\\
\vdots & \vdots & \vdots & \vdots\\
\sum b_{1} & \sum b_{2} & \cdots & \sum b_{M}
\end{array}\right]_{M\times M}W.
\end{alignat*}
Since $\left(1\right)$ is downward convex and is zero when $\alpha=0$, the extra term can take a negative value.
Then, the optimal $\alpha$ is
\begin{alignat*}{1}
FOC: & 0=2n\alpha W^{\top}\left[\begin{array}{cccc}
1 & 1 & \cdots & 1\\
1 & \ddots &  & \vdots\\
\vdots\\
1 & \cdots &  & 1
\end{array}\right]_{M\times M}W-W^{\top}\left[\begin{array}{cccc}
\sum b_{1} & \sum b_{1} & \cdots & \sum b_{1}\\
\sum b_{2} & \sum b_{2} & \cdots & \sum b_{2}\\
\vdots & \vdots & \vdots & \vdots\\
\sum b_{M} & \sum b_{M} & \cdots & \sum b_{M}
\end{array}\right]_{M\times M}W\\
 & -W^{\top}\left[\begin{array}{cccc}
\sum b_{1} & \sum b_{2} & \cdots & \sum b_{M}\\
\sum b_{1} & \sum b_{2} & \cdots & \sum b_{M}\\
\vdots & \vdots & \vdots & \vdots\\
\sum b_{1} & \sum b_{2} & \cdots & \sum b_{M}
\end{array}\right]_{M\times M}W\\
\alpha & =\frac{W^{\top}\left\{ \left[\begin{array}{cccc}
\sum b_{1} & \sum b_{1} & \cdots & \sum b_{1}\\
\sum b_{2} & \sum b_{2} & \cdots & \sum b_{2}\\
\vdots & \vdots & \vdots & \vdots\\
\sum b_{M} & \sum b_{M} & \cdots & \sum b_{M}
\end{array}\right]_{M\times M}+\left[\begin{array}{cccc}
\sum b_{1} & \sum b_{2} & \cdots & \sum b_{M}\\
\sum b_{1} & \sum b_{2} & \cdots & \sum b_{M}\\
\vdots & \vdots & \vdots & \vdots\\
\sum b_{1} & \sum b_{2} & \cdots & \sum b_{M}
\end{array}\right]_{M\times M}\right\} W}{2nW^{\top}\left[\begin{array}{cccc}
1 & 1 & \cdots & 1\\
1 & \ddots &  & \vdots\\
\vdots\\
1 & \cdots &  & 1
\end{array}\right]_{M\times M}W}.
\end{alignat*}

Adding the mean-shift $\alpha$ to linear averaging, the component for location bias can be written as
\[
\alpha_{M}=\sum_{h=1}^{k_{M}}w_{h}^{*}\hat{\alpha}_{h}+\alpha,
\]
which uniformly improves the total error by fully controlling for location bias without compromising minimizing regression error.
Additionally, adding $\alpha$ changes the optimal solution with regard to the weights $w$, which possibly improves the regression error as well.

\subsection{Criterion}
Since the loss is defined as
\begin{align*}
L\left(W,\alpha\right) & =\left(\mu-\hat{\mu}\left(W\right)-\alpha\right)^{\top}\left(\mu-\hat{\mu}\left(W\right)-\alpha\right)\\
 & =\left(\mu-\mu\left(W\right)-\alpha\right)^{\top}\left(\mu-\mu\left(W\right)-\alpha\right)\\
 & +2\alpha1^{\top}\left(\mu\left(W\right)-\hat{\mu}\left(W\right)\right)+\left(\mu\left(W\right)-\hat{\mu}\left(W\right)\right)^{\top}\left(\mu\left(W\right)-\hat{\mu}\left(W\right)\right),
 \end{align*}
 the expectation is
 \begin{align*}
\mathbb{E}\left[L_{n}\left(W\right)\left|X\right.\right] & =\left(\left(I-P\left(W\right)\right)\mu-\alpha1\right)^{\top}\left(\left(I-P\left(W\right)\right)\mu-\alpha1\right)+W^{\top}\sigma\varGamma_{M}W.
\end{align*}
Then,  the mean-shift Mallows criterion is
\begin{align*}
C_{n}\left(W,\alpha\right)&=  \left(y-\hat{\mu}\left(W\right)-\alpha1\right)^{\top}\left(y-\hat{\mu}\left(W\right)-\alpha1\right)+2\sigma^{2}K^{\top}W\\
&=  \left(\mu-\hat{\mu}\left(W\right)-\alpha1\right)^{\top}\left(\mu-\hat{\mu}\left(W\right)-\alpha1\right)+2\sigma^{2}K^{\top}W\\
 & +\varepsilon^{\top}\varepsilon-2\varepsilon^{\top}\left(\mu-\hat{\mu}\left(W\right)-\alpha1\right)\\
K & =\left(k_{1},k_{2},\cdots,k_{m}\right)^{\top}\\
\mathbb{E}\left[C_{n}\left(W,\alpha\right)\right] & =\mathbb{E}\left[L_{n}\left(W,\alpha\right)\right]+n\sigma^{2},
\end{align*}
which can be written as
\begin{align*}
C_{n}\left(W,\alpha\right)= & W^{\top}\left(y-x\hat{\beta}-\alpha1\right)^{\top}\left(y-x\hat{\beta}-\alpha1\right)W+2\sigma^{2}K^{\top}W\\
= & W^{\top}\left(\mu-x\hat{\beta}-\alpha1\right)^{\top}\left(\mu-x\hat{\beta}-\alpha1\right)W+2\sigma^{2}K^{\top}W\\
 & +W^{\top}\varepsilon^{\top}\varepsilon W+2W^{\top}\varepsilon^{\top}\left(\mu-x\hat{\beta}-\alpha1\right)W.
\end{align*}

Comparing the criterion with and without an intercept, there exists an $\alpha$ that satisfies
\[
\min_{W,\alpha}C_{n}\left(W,\alpha\right)\leqq\min_{W}C_{n}\left(W\right).
\]
This can be seen by setting $\alpha=0$ with regard to $\left(W,\alpha\right)$ that satisfies $C_{n}\left(W\right)\leqq C_{n}\left(W,\alpha\right)$.

\subsection{Optimality}
Given the extension, we now show that our proposed MSA is optimal to the MMA estimator in \cite{hansen2007least}.
We set the parameter space $K$ for which the intercept can take values,
\[
K_{n}=\left\{ \alpha\left|\min_{W}C_{n}\left(W,\alpha\right)\leqq\min_{W}C_{n}\left(W\right)\right.\right\}.
\]
Here, $\left\{ K_{n}\right\} $ is on $\mathbb{R}$ including zero.
Additionally, we assume
\[
\mathbb{E}\left[\left|e_{i}\right|^{4\left(N+1\right)}\left|x_{i}\right.\right]\leqq\kappa<\infty.
\]

When $n\rightarrow\infty$, we assume,
\[
\xi_{n}\triangleq{\displaystyle \inf_{\begin{array}{c}
W\in H_{n}\left(W\right)\\
\alpha\in K_{n}
\end{array}}}R_{n}\left(W,\alpha\right)\rightarrow\infty,
\]
almost surely.
We additionally assume, for each point, $w\in H_{n}$,
\begin{equation}
\frac{R_{n}\left(w,\alpha\right)}{\xi_{n}^{2}}\rightarrow0,\ ^{\forall}w\in H_{n}\label{eq:A3}.
\end{equation}
This condition is weaker than that of \cite{wan2010least}.
Compared to the conditions in \cite{hansen2007least},
\[
\inf_{w\in H_{n}}R_{n}\left(w\right)\rightarrow\infty,
\]
this might seem somewhat restrictive.
However, we only need to add the following condition,
\[
\frac{R_{n}\left(w,\alpha\right)}{{\displaystyle \inf_{\begin{array}{c}
W\in H_{n}\left(W\right)\\
\alpha\in K_{n}
\end{array}}}R_{n}\left(w,\alpha\right)}<\infty,^{\forall}w\in H_{n}
\]
to the above, which is a reasonable condition to add.

\begin{proposition}
{\it Given the above assumptions, we have,}
\[
\frac{L_{n}\left(\hat{W}_{N},\hat{\alpha}\right)}{{\displaystyle \inf_{\begin{array}{c}
W\in H_{n}\left(W\right)\\
\alpha\in K_{n}
\end{array}}}L_{n}\left(W,\alpha\right)}\rightarrow1.
\]
\end{proposition}

The proof roughly follows that of \cite{li1987asymptotic} and \cite{hansen2007least}, though the loss, risk, and criterion are dependent on the continuous parameter, $\left(W,\alpha\right)$, which makes the proof challenging.
Note that \cite{wan2010least} provides the conditions to relax the  discreteness of the weights.
In contrast, we prove optimality using the convexity lemma by utilizing the fact that the loss and criterion is a convex function.
This allows us to prove optimality with a weaker condition to that of \cite{wan2010least}.

We first decompose $C_{n}\left(W,\alpha\right)$ as
\begin{alignat*}{1}
C_{n}\left(W,\alpha\right)= & \left(y-\hat{\mu}\left(W\right)-\alpha1\right)^{\top}\left(y-\hat{\mu}\left(W\right)-\alpha1\right)+2\sigma^{2}K^{\top}W\\
= & \left(\mu-\hat{\mu}\left(W\right)-\alpha1\right)^{\top}\left(\mu-\hat{\mu}\left(W\right)-\alpha1\right)\\
 & +2\sigma^{2}K^{\top}W-2\varepsilon^{\top}\left(\mu-\hat{\mu}\left(W\right)-\alpha1\right)\\
 & +\varepsilon^{\top}\varepsilon\\
= & L_{n}\left(W,\alpha\right)+2\left\langle \varepsilon,A_{m}\mu-\alpha1\right\rangle \\
 & +2\sigma^{2}K^{\top}W-2\left\langle \varepsilon,P\left(W\right)\varepsilon\right\rangle +\varepsilon^{\top}\varepsilon.
\end{alignat*}
Since the $\varepsilon^{\top}\varepsilon$ term does not affect model selection, this is equivalent to minimizing,
\[
L_{n}\left(W,\alpha\right)+2\left\langle \varepsilon,A_{m}\mu-\alpha1\right\rangle +2\sigma^{2}K^{\top}W-2\left\langle \varepsilon,P\left(W\right)\varepsilon\right\rangle.
\]

To prove optimality, we must show
\[
{\displaystyle \min_{W,\alpha}}\frac{C_{n}\left(W,\alpha\right)}{{\displaystyle \inf_{\begin{array}{c}
W\in H_{n}\left(W\right)\\
\alpha\in K_{n}
\end{array}}}L_{n}\left(W,\alpha\right)}\rightarrow1,
\]
which implies that
\[
P\left(\sup_{\begin{array}{c}
W\in H_{n}\left(W\right)\\
\alpha\in K_{n}
\end{array}}\left|\frac{C_{p}\left(h\right)}{{\displaystyle \inf_{\begin{array}{c}
W\in H_{n}\left(W\right)\\
\alpha\in K_{n}
\end{array}}}L_{n}\left(W,\alpha\right)}-1\right|>\delta\right)\rightarrow0
\]
must uniformly hold with regard to $W\in H_{n}$ and $\alpha\in K_{n}$.

Since $\frac{C_{n}\left(W,\alpha\right)}{L_{n}\left(W,\alpha\right)}$ can be written as
\[
\frac{L_{n}\left(W,\alpha\right)+2\left\langle \varepsilon,A_{m}\mu-\alpha1\right\rangle +\left\{ 2\sigma^{2}K^{\top}W-2\left\langle \varepsilon,P\left(W\right)\varepsilon\right\rangle \right\} }{R_{n}\left(W,\alpha\right)}\frac{R_{n}\left(W,\alpha\right)}{L_{n}\left(h\right)}
\]
and each term is
\begin{alignat*}{1}
\left|\frac{2\left\langle \varepsilon,A_{m}\mu-\alpha1\right\rangle }{R_{n}\left(W,\alpha\right)}\right| & \leqq\left|\frac{2\left\langle \varepsilon,A_{m}\mu-\alpha1\right\rangle }{\xi_{n}}\right|\\
\frac{\left|2\sigma^{2}K^{\top}W-2\left\langle \varepsilon,P\left(W\right)\varepsilon\right\rangle \right|}{R_{n}\left(W,\alpha\right)} & \leqq\frac{\left|2\sigma^{2}K^{\top}W-2\left\langle \varepsilon,P\left(W\right)\varepsilon\right\rangle \right|}{\xi_{n}}\\
\left|\frac{L_{n}\left(W,\alpha\right)}{R_{n}\left(W,\alpha\right)}-1\right| & \leqq\frac{\left|L_{n}\left(W,\alpha\right)-R_{n}\left(W,\alpha\right)\right|}{\xi_{n}},
\end{alignat*}
we need to show the following to prove optimality:
\begin{description}
\item [{2.2}] 
\[
\sup_{\begin{array}{c}
W\in H_{n}\left(W\right)\\
\alpha\in K_{n}
\end{array}}\left|\frac{2\left\langle \varepsilon,A_{m}\mu-\alpha1\right\rangle }{\xi_{n}}\right|\rightarrow0,
\]
\item [{2.3}] 
\[
\sup_{\begin{array}{c}
W\in H_{n}\left(W\right)\\
\alpha\in K_{n}
\end{array}}\frac{\left|2\sigma^{2}K^{\top}W-2\left\langle \varepsilon,P\left(W\right)\varepsilon\right\rangle \right|}{\xi_{n}}\rightarrow0,
\]
\item [{2.4}] 
\[
\sup_{\begin{array}{c}
W\in H_{n}\left(W\right)\\
\alpha\in K_{n}
\end{array}}\frac{\left|L_{n}\left(W,\alpha\right)-R_{n}\left(W,\alpha\right)\right|}{\xi_{n}}\rightarrow0.
\]
\end{description}

To show uniform convergence, it is sufficient to show pointwise convergence, following the below convexity lemma:
\begin{theorem}
\citep{rockafellar1970convex,pollard1991asymptotics}. {\it Consider the real convex function $\left\{ \rho_{n}\left(\theta\right)\left|\theta\in\Theta\right.\right\} $ with regard to $\theta$, then $\Theta\subset\mathbb{R}^{d}$ is a convex open set.
If $\rho_{n}\left(\theta\right)$ pointwise converges to $\rho$, then $\rho_{n}\left(\theta\right)$ compact converges on $\Theta$:
\[
\sup_{\theta\in K\subset\Theta}\left|\rho_{n}\left(\theta\right)-\rho\left(\theta\right)\right|\rightarrow0,\ prob.
\]
Note, it is necessary that $\rho$ is a convex function.}
\end{theorem}

{\it Proof of} {\bf 2.2:} From Chebyshev's inequality, we have 
\[
P\left(\left|\frac{2\left\langle \varepsilon,A_{m}\mu-\alpha1\right\rangle }{\xi_{n}}\right|>\delta\right)\leqq\frac{4\mathbb{E}\left[\left\langle \varepsilon,A_{m}\mu-\alpha1\right\rangle ^{2}\right]}{\delta^{2}\xi_{n}^{2}},
\]
which we evaluate the right hand side. 
From the inequality in \cite{whittle1960bounds}, the following
\[
\mathbb{E}\left[\left\langle \varepsilon,A_{m}\mu-\alpha1\right\rangle ^{2}\right]\leqq2^{2}C\left(s\right)\sigma\left\Vert A_{m}\mu-\alpha1\right\Vert ^{2}
\]
holds, thus,
\[
\frac{4\mathbb{E}\left[\left\langle \varepsilon,A_{m}\mu-\alpha1\right\rangle ^{2}\right]}{\delta^{2}\xi_{n}^{2}}\leqq C\frac{1}{\delta^{2}}\frac{\left\Vert A_{m}\mu-\alpha1\right\Vert ^{2}}{\xi_{n}^{2}},
\]
where $C$ is some constant.
Now, since 
\[
\left\Vert A_{m}\mu-\alpha1\right\Vert ^{2}\leqq R_{n}\left(W,\alpha\right),
\]
we have 
\[
C\frac{1}{\delta^{2}}\frac{\left\Vert A_{m}\mu-\alpha1\right\Vert ^{2}}{\xi_{n}^{2}}\leqq C\frac{1}{\delta^{2}}\frac{R_{n}\left(W,\alpha\right)}{\xi_{n}^{2}}.
\]
Therefore, from assumption~\ref{eq:A3}, we prove {\bf 2.2}.

{\it Proof of} {\bf 2.3:} Similarly to {\bf 2.2}, applying  Chebyshev's inequality, we evaluate the moment of the right hand side.
Applying the quadric version of Whittle's inequality, we have 
\[
\mathbb{E}\left[\left|2\sigma^{2}K^{\top}W-2\left\langle \varepsilon,P\left(W\right)\varepsilon\right\rangle \right|^{2}\right]\leqq C\left(\sigma^{2}\textrm{tr}\left\{ P\left(W\right)P\left(W\right)\right\} \right),
\]
where, $C$ is a constant.
Additionally, since 
\[
\sigma^{2}\textrm{tr}\left\{ P\left(W\right)P\left(W\right)\right\} \leqq R_{n}\left(W,\alpha\right),
\]
we prove {\bf 2.3}.

{\it Proof of} {\bf 2.4:} For this, it is sufficient to show convergence for the following:
\begin{alignat*}{1}
\frac{\left|L_{n}\left(W,\alpha\right)-R_{n}\left(W,\alpha\right)\right|}{\xi_{n}} & =\frac{\left|L_{n}\left(W,\alpha\right)-R_{n}\left(W,\alpha\right)\right|}{\xi_{n}}\\
L_{n}\left(h\right) & =\left\Vert \mu-\mu\left(W\right)-\alpha1\right\Vert ^{2}+2\left\langle \mu-\alpha1^{\top},P\left(W\right)\varepsilon\right\rangle +\left\Vert P\left(W\right)\varepsilon\right\Vert ^{2}\\
R_{n}\left(h\right) & =\left\Vert \mu-\mu\left(W\right)-\alpha1\right\Vert ^{2}+\sigma^{2}\textrm{tr}\left\{ P\left(W\right)P\left(W\right)\right\},
\end{alignat*}
thus,
\begin{alignat*}{1}
\sup_{\left(W,\alpha\right)\in H_{n}}\frac{\left|\left\langle \mu-\alpha1^{\top},P\left(W\right)\varepsilon\right\rangle \right|}{\xi_{n}} & \rightarrow0\\
\sup_{\left(W,\alpha\right)\in H_{n}}\frac{\left|\left\Vert P\left(W\right)\varepsilon\right\Vert ^{2}-\sigma^{2}\textrm{tr}\left\{ P\left(W\right)P\left(W\right)\right\} \right|}{\xi_{n}} & \rightarrow0.
\end{alignat*}

Noting that 
\begin{alignat*}{1}
\left\langle \mu-\alpha1^{\top},P\left(W\right)\varepsilon\right\rangle  & =\left\langle P\left(W\right)\left(\mu-\alpha1^{\top}\right),\varepsilon\right\rangle \\
\left\Vert P\left(W\right)\left(\mu-\alpha1^{\top}\right)\right\Vert ^{2} & \leqq\lambda\left\{ P\left(W\right)\right\} ^{2}\left\Vert \mu-\alpha1^{\top}\right\Vert ^{2},
\end{alignat*}
the first equation is equivalent to {\bf 2.2}, and the second equation, utilizing the quadratic version of the Whittle's inequality to evaluate the expectation,
\[
\mathbb{E}\left[\left|\left\Vert P\left(W\right)\varepsilon\right\Vert ^{2}-\sigma^{2}\textrm{tr}\left\{ P\left(W\right)P\left(W\right)\right\} \right|^{2}\right]
\]
and further noting that 
\begin{alignat*}{1}
\left\Vert P\left(W\right)\varepsilon\right\Vert ^{2} & =\left\langle P\left(W\right)P\left(W\right)\varepsilon,\varepsilon\right\rangle \\
\textrm{tr}\left\{ P\left(W\right)P\left(W\right)\right\} ^{2} & \leqq\lambda\left\{ P\left(W\right)\right\} ^{2}\textrm{tr}\left\{ P\left(W\right)P\left(W\right)\right\},
\end{alignat*}
for some constant, $C$, we have the following:
\[
\mathbb{E}\left[\left|\left\Vert P\left(W\right)\varepsilon\right\Vert ^{2}-\sigma^{2}\textrm{tr}\left\{ P\left(W\right)P\left(W\right)\right\} \right|^{2}\right]\leqq CR_{n}\left(W,\alpha\right).
\]
The rest is the same as {\bf 2.3}. Q.E.D.

\section{Finite sample investigation \label{sec:sim}}
We investigate the finite sample mean squared error of our proposed MSA estimator to that of the original MMA estimator in a simple simulation setting similar to \cite{hansen2007least}.
Our data generating process is a slight modification of eq.~\eqref{dgp}, where the misspecification bias is positive:
\begin{align*}
y_{i}=\sum_{j=1}^{k_{m}}\theta_{j}x_{ji}+\sum_{j=k_{m}+1}^{\infty}\theta_{j}\textrm{exp}(x_{ji})+\varepsilon_{i},
\end{align*}
where all $x_{ji}$ are independent and identically distributed $N(0,1)$, the error, $\varepsilon_{i}$, is $N(0,1)$ and independent of $x_{ji}$.
The parameters are determined by the rule, $\theta_{j}=c\sqrt{2\alpha}j^{-\alpha -1/2}$, where $c=R^2/(1-R^2)$.
Though not reported, the results are robust to alternative specifications, except when the bias has expectation zero (i.e. no location bias), at which MSA is equivalent to MMA.

For the finite sample investigation, we vary the sample size by $n=50,150,400, \textrm{ and } 1,000$ and $\alpha=0.5,1, \textrm{ and } 1.5$.
The parameter $\alpha$ controls the rate of decline of the coefficients, $\theta_{j}$, where a larger $\alpha$ implies a steeper decline in $j$.
The number of models is determined by $M=3n^{1/3}$, which, for our four sample sizes, means $M=11,16,22,\textrm{ and } 30$.
We vary $c$ so as $R^2$ varies from 0.1 to 0.9.

Unlike \cite{hansen2007least}, which compared the MMA estimator to AIC model selection, Mallows model selection, smoothed AIC, and smoothed BIC, we simply compare our MSA estimator to the MMA estimator, as \cite{hansen2007least} convincingly show that the MMA estimator is superior to these methods.
To evaluate the two estimators, we compute the log risk (expected squared error) for each and calculate the difference.
The simulation is done for 1,000 iterations and averaged over.

\begin{figure}[htbp!]
\centering
\includegraphics[width=0.9\textwidth]{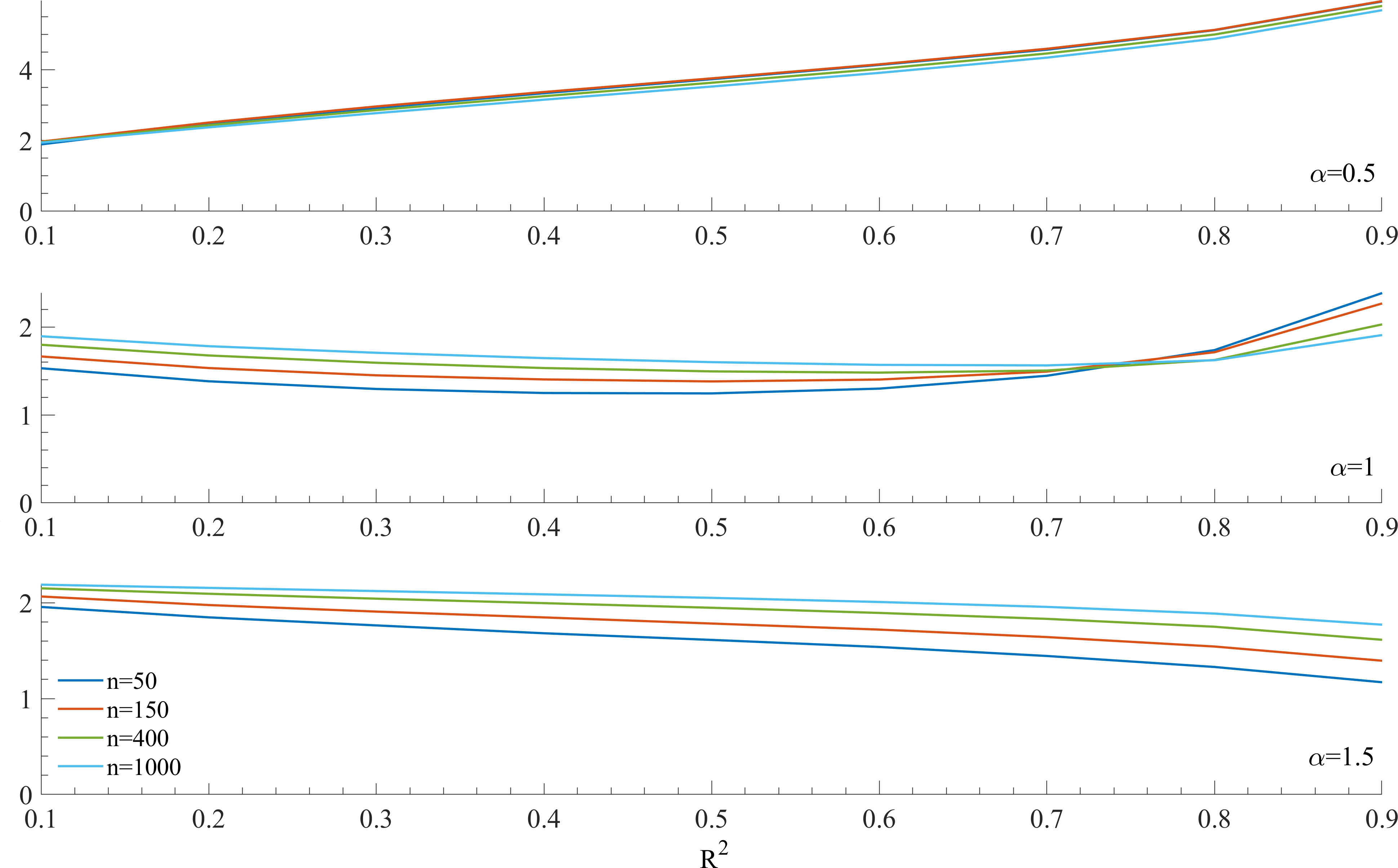} 
\caption{Difference in log risk between MMA and MSA for $n=50,150,400, \textrm{ and } 1,000$ samples and $\alpha=0.5,1, \textrm{ and } 1.5$. Risk is calculated as the average over 1,000 iterations.
\label{fig:sim}}
\end{figure}

The results of the finite sample investigation are shown in Fig.~\ref{fig:sim}.
The three panels display the difference in log risk between MMA and MSA for the varying $\alpha$, where the difference is displayed on the $y$ axis and the population $R^2$ is displayed on the $x$ axis.
The different colored lines are for the different sample sizes, $n$.

For all results, our proposed MSA estimator uniformly outperforms the MMA estimator to varying degrees.
In the case of $\alpha=0.5$, the results are almost identical for all sample sizes.
The relative performance of MSA increases as $R^2$ increases, which is expected as the increased $R^2$, under a low $\alpha$, implies a more robust bias component.
This bias component is captured in MSA, leading to improved performances.
As $\alpha$ increases, the results invert, in the sense that, for $\alpha=1$, the difference in log risk is bowed, while, for $\alpha=1.5$, the difference decreases as $R^2$ increases.
This is not unexpected as, in \cite{hansen2007least}, the performance of the MMA estimator decreases in these cases.
If anything, for $\alpha=1$, the MSA estimator seems more resilient for higher $R^2$.
Over all variations, the MSA estimator is superior to the MMA estimator.

\bibliographystyle{elsarticle-harv}
\bibliography{reference}

\begin{thebibliography}{35}
\expandafter\ifx\csname natexlab\endcsname\relax\def\natexlab#1{#1}\fi
\expandafter\ifx\csname url\endcsname\relax
  \def\url#1{\texttt{#1}}\fi
\expandafter\ifx\csname urlprefix\endcsname\relax\def\urlprefix{URL }\fi

\bibitem[{Aastveit et~al.(2014)Aastveit, Gerdrup, Jore, and
  Thorsrud}]{Aastveit2014}
Aastveit, K.~A., Gerdrup, K.~R., Jore, A.~S., Thorsrud, L.~A., 2014. Nowcasting
  {GDP} in real time: {A} density combination approach. Journal of Business \&
  Economic Statistics 32, 48--68.

\bibitem[{Aastveit et~al.(2018{\natexlab{a}})Aastveit, Mitchell, Ravazzolo, and
  van Dijk}]{aastveit2018evolution}
Aastveit, K.~A., Mitchell, J., Ravazzolo, F., van Dijk, H.~K.,
  2018{\natexlab{a}}. The evolution of forecast density combinations in
  economics. Tech. rep., Tinbergen Institute Discussion Paper.

\bibitem[{Aastveit et~al.(2018{\natexlab{b}})Aastveit, Ravazzolo, and
  Van~Dijk}]{Aastveit2015}
Aastveit, K.~A., Ravazzolo, F., Van~Dijk, H.~K., 2018{\natexlab{b}}. Combined
  density nowcasting in an uncertain economic environment. Journal of Business
  \& Economic Statistics 36~(1), 131--145.

\bibitem[{Amisano and Giacomini(2007)}]{Amisano2007}
Amisano, G.~G., Giacomini, R., 2007. Comparing density forecasts via weighted
  likelihood ratio tests. Journal of Business \& Economic Statistics 25,
  177--190.

\bibitem[{Bates and Granger(1969)}]{BatesGranger1969}
Bates, J.~M., Granger, C. W.~J., 1969. The combination of forecasts.
  Operational Research Quarterly 20, 451--468.

\bibitem[{Bernardo and Smith(2009)}]{bernardo2009bayesian}
Bernardo, J.~M., Smith, A.~F., 2009. Bayesian theory. Vol. 405. John Wiley \&
  Sons.

\bibitem[{Billio et~al.(2012)Billio, Casarin, Ravazzolo, and van
  Dijk}]{Billio2012}
Billio, M., Casarin, R., Ravazzolo, F., van Dijk, H.~K., 2012. Combination
  schemes for turning point predictions. Quarterly Review of Finance and
  Economics 52, 402--412.

\bibitem[{Billio et~al.(2013)Billio, Casarin, Ravazzolo, and van
  Dijk}]{Billio2013}
Billio, M., Casarin, R., Ravazzolo, F., van Dijk, H.~K., 2013. Time-varying
  combinations of predictive densities using nonlinear filtering. Journal of
  Econometrics 177, 213--232.

\bibitem[{Breiman(1996)}]{breiman1996bagging}
Breiman, L., 1996. Bagging predictors. Machine learning 24~(2), 123--140.

\bibitem[{D{\v{z}}eroski and {\v{Z}}enko(2004)}]{dvzeroski2004combining}
D{\v{z}}eroski, S., {\v{Z}}enko, B., 2004. Is combining classifiers with
  stacking better than selecting the best one? Machine learning 54~(3),
  255--273.

\bibitem[{Geweke and Amisano(2012)}]{Geweke2012}
Geweke, J., Amisano, G.~G., 2012. Prediction with misspecified models. The
  American Economic Review 102, 482--486.

\bibitem[{Geweke and Amisano(2011)}]{Geweke2011}
Geweke, J.~F., Amisano, G.~G., 2011. Optimal prediction pools. Journal of
  Econometrics 164, 130--141.

\bibitem[{Hall and Mitchell(2007)}]{HallMitchell2007}
Hall, S.~G., Mitchell, J., 2007. Combining density forecasts. International
  Journal of Forecasting 23, 1--13.

\bibitem[{Hansen(2007)}]{hansen2007least}
Hansen, B.~E., 2007. Least squares model averaging. Econometrica 75~(4),
  1175--1189.

\bibitem[{Hansen and Racine(2012)}]{hansen2012jackknife}
Hansen, B.~E., Racine, J.~S., 2012. Jackknife model averaging. Journal of
  Econometrics 167~(1), 38--46.

\bibitem[{Hoeting et~al.(1999)Hoeting, Madigan, Raftery, and
  Volinsky}]{hoeting1999bayesian}
Hoeting, J.~A., Madigan, D., Raftery, A.~E., Volinsky, C.~T., 1999. Bayesian
  model averaging: a tutorial. Statistical science, 382--401.

\bibitem[{Hoogerheide et~al.(2010)Hoogerheide, Kleijn, Ravazzolo, Van~Dijk, and
  Verbeek}]{Hooger2010}
Hoogerheide, L., Kleijn, R., Ravazzolo, F., Van~Dijk, H.~K., Verbeek, M., 2010.
  Forecast accuracy and economic gains from {B}ayesian model averaging using
  time-varying weights. Journal of Forecasting 29, 251--269.

\bibitem[{Kapetanios et~al.(2015)Kapetanios, Mitchell, Price, and
  Fawcett}]{Fawcett2014}
Kapetanios, G., Mitchell, J., Price, S., Fawcett, N., 2015. Generalised density
  forecast combinations. Journal of Econometrics 188, 150--165.

\bibitem[{Kascha and Ravazzolo(2010)}]{Kascha2010}
Kascha, C., Ravazzolo, F., 2010. Combining inflation density forecasts. Journal
  of Forecasting 29, 231--250.

\bibitem[{Li et~al.(1987)}]{li1987asymptotic}
Li, K.-C., et~al., 1987. Asymptotic optimality for $ c\_p, c\_l $,
  cross-validation and generalized cross-validation: Discrete index set. The
  Annals of Statistics 15~(3), 958--975.

\bibitem[{Louppe(2014)}]{louppe2014understanding}
Louppe, G., 2014. Understanding random forests: From theory to practice. arXiv
  preprint arXiv:1407.7502.

\bibitem[{McAlinn et~al.(2019)McAlinn, Aastveit, Nakajima, and
  West}]{mcalinn2017multivariate}
McAlinn, K., Aastveit, K.~A., Nakajima, J., West, M., 2019. Multivariate
  bayesian predictive synthesis in macroeconomic forecasting. Journal of the
  American Statistical Association forthcoming.

\bibitem[{McAlinn and West(2019)}]{mcalinn2019dynamic}
McAlinn, K., West, M., 2019. Dynamic bayesian predictive synthesis in time
  series forecasting. Journal of Econometrics 210~(1), 155--169.

\bibitem[{Negro et~al.(2016)Negro, Hasegawa, and Schorfheid}]{Negro2016}
Negro, M.~D., Hasegawa, R.~B., Schorfheid, F., 2016. Dynamic prediction pools:
  an investigation of financial frictions and forecasting performance. Journal
  of Econometrics 192(2), 391--405.

\bibitem[{Pettenuzzo and Ravazzolo(2016)}]{Pettenuzzo2015}
Pettenuzzo, D., Ravazzolo, F., 2016. Optimal portfolio choice under
  decision-based model combinations. Journal of Applied Econometrics 31~(7),
  1312--1332.

\bibitem[{Pollard(1991)}]{pollard1991asymptotics}
Pollard, D., 1991. Asymptotics for least absolute deviation regression
  estimators. Econometric Theory 7~(2), 186--199.

\bibitem[{Raftery et~al.(1997)Raftery, Madigan, and
  Hoeting}]{raftery1997bayesian}
Raftery, A.~E., Madigan, D., Hoeting, J.~A., 1997. Bayesian model averaging for
  linear regression models. Journal of the American Statistical Association
  92~(437), 179--191.

\bibitem[{Rockafellar(1970)}]{rockafellar1970convex}
Rockafellar, R.~T., 1970. Convex analysis. Vol.~28. Princeton university press.

\bibitem[{Schapire(2003)}]{schapire2003boosting}
Schapire, R.~E., 2003. The boosting approach to machine learning: An overview.
  In: Nonlinear estimation and classification. Springer, pp. 149--171.

\bibitem[{Takanashi and McAlinn(2019)}]{takanashi2019predictive}
Takanashi, K., McAlinn, K., 2019. Predictive properties of forecast
  combination, ensemble methods, and bayesian predictive synthesis. arXiv
  preprint arXiv:1911.08662.

\bibitem[{Terui and van Dijk(2002)}]{Terui2002}
Terui, N., van Dijk, H.~K., 2002. Combined forecasts from linear and non-
  linear time series models. International Journal of Forecasting 18(3),
  421--438.

\bibitem[{Wan et~al.(2010)Wan, Zhang, and Zou}]{wan2010least}
Wan, A.~T., Zhang, X., Zou, G., 2010. Least squares model averaging by mallows
  criterion. Journal of Econometrics 156~(2), 277--283.

\bibitem[{Whittle(1960)}]{whittle1960bounds}
Whittle, P., 1960. Bounds for the moments of linear and quadratic forms in
  independent variables. Theory of Probability \& Its Applications 5~(3),
  302--305.

\bibitem[{Yao et~al.(2018)Yao, Vehtari, Simpson, Gelman, et~al.}]{yao2018using}
Yao, Y., Vehtari, A., Simpson, D., Gelman, A., et~al., 2018. Using stacking to
  average bayesian predictive distributions (with discussion). Bayesian
  Analysis 13~(3), 917--1003.

\bibitem[{Zhang et~al.(2013)Zhang, Wan, and Zou}]{zhang2013model}
Zhang, X., Wan, A.~T., Zou, G., 2013. Model averaging by jackknife criterion in
  models with dependent data. Journal of Econometrics 174~(2), 82--94.

\end{thebibliography}
\end{document}